\documentclass[twocolumn,showpacs,preprintnumbers,amsmath,amssymb]{revtex4}

\usepackage{graphicx}
\usepackage{dcolumn}
\usepackage{bm}

\begin{document}


\title{Polaron with disordered electron-phonon interaction}

\author{B.Ya.Yavidov}
\affiliation{%
Nukus State Pedagogical Institute named after A'jiniyaz, 230105
Nukus, Uzbekistan
}%

\date{\today}

\begin{abstract}
A single electron in one dimensional lattice is considered within
the framework of extended Holstein model at strong-coupling limit.
Disordered density-displacement type electron-phonon interaction is
proposed. Basic parameters of small polaron formed due to disordered
electron-phonon interaction are calculated. It is shown that
disordered electron-phonon interaction substantially influences all
properties of the polaron. Depending on disordered electron-phonon
interaction polaronic effect might be enhanced or diminished. It
turns out that all parameters of the polaron are site dependent.
Further application of proposed model is briefly discussed.
\end{abstract}

\maketitle

\section{Introduction}

Polaron concept was introduced in the 30$^{th}$ years of XX century
by Landau \cite{lan} and later was theoretically developed by a
number of researchers (see  review \cite{asa-devr-adv-pol-phys}).
Though in the field of polaron physics there are substantial
understanding of main features of polaron related phenomena in a
variety of solids, there still remain an open issues to be
clarified. One of them is the application of the theoretical models
to the real solids, in particular the various disordered systems. An
interplay of lattice disorder and electron-phonon interaction in
such systems, as it is well known, determines charge carrier
dynamics. In the past decades Holstein polarons \cite{hol} were
studied in a variety of medium, including near an impurity
\cite{hague-k-asa-impurity-pol,mishchenko-nagaosa-etal,ebrahim-berciu-85}
and in a disordered lattice
\cite{cohen-economou-souk-1983,bronold-fehske-prb-66-073102,Chat-Das,berciu-sawatzky,berciu-mishchenko-nagaosa,ebrahim-berciu-86,tozer-barfold}.
The all above mentioned works except of
Ref.\cite{hague-k-asa-impurity-pol} deal with local electron-phonon
interaction, in which charge carrier interacts only with onsite
intramolecular vibrations. In reality, charge carrier interacts
simultaneously with all ions of the lattice and the range of this
electron-phonon interaction extends over many lattice units. At the
same time disorder influences coupled electron-phonon system too.
Noteworthy that in work \cite{ebrahim-berciu-85} it was shown
opposite effect namely that electron-phonon interaction renormalizes
disorder potential and that disorder seen by a polaron is different
from the bare disorder\cite{ebrahim-berciu-86}. In this sense, an
electron sees the disordered ions and as a consequence it interacts
with them via electron-phonon interaction which is spatially
disordered as well. Such type of disordered electron-phonon
interaction and its role in polaron formation was not studied yet.
The objective of this paper is to remedy this shortcoming. In order
to do this we further develop an idea of Ref.\cite{alekor} with
regard of polarons and apply it to disordered lattice.

Namely in Ref.\cite{alekor} a model of a polaron with a long-range
"density-displacement" type force was introduced. The model by
itself represents an extension of the Fr\"{o}hlich polaron model
\cite{froh} to a discrete ionic crystal lattice or extension of the
Holstein polaron model \cite{hol} to a case when an electron
interacts with many ions of a lattice with longer ranged
electron-phonon interaction. Subsequently, the model was named as
the extended Holstein model (EHM) \cite{flw}. The model was
introduced in order to mimic  $high-T_{c}$ cuprates, where the
in-plane ($CuO_{2}$) carriers are strongly coupled to the $c$-axis
polarized vibrations of the $apical$ oxygen ions \cite{timusk}. In
the last decades the model was studied in great detail in
Refs.\cite{flw,kor-ctqmc,kor-giant,kor-ground,bt,asa-kor-pha,asa-kor-jpcm,pcf-jpcm,cfmp-prb,hohen,asa-ya,stojan,spencer,meisel,hague-etal,kor-jpcm,hague-sam,hague-kor,yav-pla,chak-min-prb-85,chak-min-prb-88,tam-tsai-kemp-prb-89,chak-tez-min-prb-89}.

\section{The model}

In this paper we deal with the extended Holstein model in one
dimensional lattice in which electron-phonon interaction is
disordered. The Hamiltonian of the model  is \cite{alekor}
\begin{equation}\label{1}
H=H_{e}+H_{ph}+H_{e-ph}
\end{equation}
where
\begin{equation}\label{2}
H_{e}=-t \sum_{\bf n}(c^{\dagger}_{\bf n}c_{\bf n+a} +H.c.)
\end{equation}
is the electron hopping energy,
\begin{equation}\label{3}
H_{ph}=\sum_{{\bf m}}\left(-\frac{\hbar^2\partial^2}{2M\partial
u^{2}_{{\bf m}}}+\frac{M\omega^2u^{2}_{{\bf m}}}{2}\right)
\end{equation}
is the Hamiltonian of the vibrating ions,
\begin{equation}\label{4}
H_{e-ph}=\sum_{{\bf n,m}}f_{{\bf m}}({\bf n})\cdot u_{{\bf
m}}c^{\dagger}_{\bf n}c_{\bf n}
\end{equation}
describes interaction between the electron that belongs to a lower
chain and the ions of an upper chain (see below). Here $t$ is the
nearest neighbor hopping integral, $c^{\dagger}_{\bf n}$($c_{\bf
n}$) is a creation (destruction) operator of an electron on a cite
$\bf n$, $u_{{\bf m}}$ is the $c$- polarized displacement of the
{\bf m}-th ion and $f_{{\bf m}}({\bf n})$ is an interacting
density-displacement type force between an electron on a site {\bf
n} and the $c$- polarized vibration of the {\bf m}-th ion. $M$ is
the mass of the vibrating ions and $\omega$ is their frequency.

\begin{figure}
\resizebox{0.5\textwidth}{!}{%
  \includegraphics{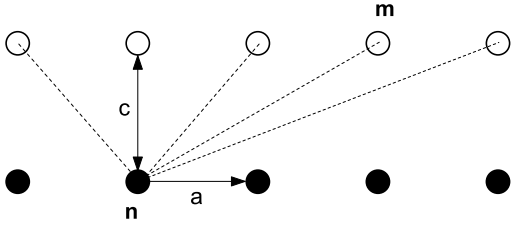}
}
\caption{An electron hops on a lower chain and interacts with the
ions vibrations of an upper infinite chain via a
density-displacement type force $f_{{\bf m}}(\bf n)$. The distances
between the chains ($|{\bf c}|$) and between the ions ($|{\bf a}|$)
are assumed equal to 1. Dotted lines represent the interaction of an
electron on site ${\bf n}$ with the ions of the upper chain.}
\label{fig.1}       
\end{figure}

We consider an electron performing hopping motion on a lower chain
consisting of the static sites, but interacting with all ions of an
upper chain via a long-range density-displacement type force, as
shown in Fig.\ref{fig.1}. So, the motion of an electron is always
one-dimensional, but a vibration of the upper chain's ions is $c$-
polarized (perpendicular to the chains). There is no doubt that an
explicit analytical form of the force $f_{\bf {m}}({\bf n})$ is one
of crucial aspects determining polaron parameters. Of cause, it
depends on structural elements that located on sites $\bf m$.
Whether the structural elements are neutral, charged (positively or
negatively) or dipoles (electrical or magnet) this force may have
different origin and may lead to different polaronic states. As in
Refs.\cite{kor-giant,spencer,hague-etal,hague-kor} here it is also
assumed that the structural elements are electrically charged
(positive or negative) and thus the force has Coulombic nature. Now
let's try to simulate  a disorder in the lattice. There are many
ways to simulate a disorder in the lattice. One of them is
allocation driven disorder where positions of upper chain's ions are
randomized. Of course, if allocation of the ions throughout the
upper chain is disordered then density-displacement type force
$f_{{\bf m}}(\bf n)$ is also disordered. An another way to simulate
a disorder in the lattice is randomize the charge value of the upper
chain's ions. There are a lot of examples for the systems where such
type of charge fluctuations or modulations occurs (see for example
\cite{ohtomo-nat-419,peng-etal-prb-62,hashimoto-etal}). Modulations
of charge states of structural elements of the lattice may occur in
many other systems like cuprates, manganites, organic semiconductors
and it has become possible owing to modern crystal growth
technology. Here we consider only charge modulation (fluctuation)
driven disorder of electron-phonon interaction and its consequences
in polaron formation. So, our consideration is relevant to all
compounds where polaron formation is possible and it occurs in the
presence of charge modulation's (fluctuation) driven electron-phonon
interaction.

\begin{figure}
\resizebox{0.5\textwidth}{!}{%
  \includegraphics{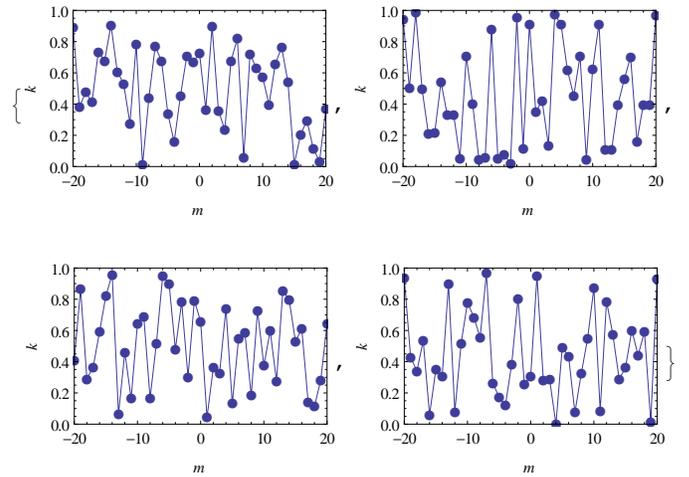}
}
\caption{The value of $k$ at $m$ for different types of disorder of
upper chain's ions. $k$ is taken as uniformly distributed randomized
number in the interval (0,1).}
\label{fig.2}       
\end{figure}

\begin{figure}
\resizebox{0.5\textwidth}{!}{%
  \includegraphics{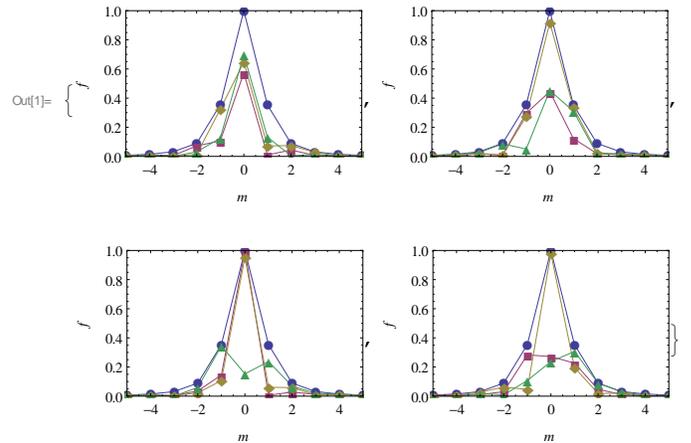}
}
\caption{Dependence of {\it disordered} density-displacement type
electron-phonon interaction force $f_{\bf{m}}(\bf{0})$ on $\bf{m}$
for different types of disorder of upper chain.}
\label{fig.3}       
\end{figure}

Considering the above said opinions without loss of generality one
writes explicitly a discrete form of the density-displacement type
electron-phonon interaction force as:

\begin{equation}\label{5}
f_{{\bf m}}({\bf n})=\frac{c\cdot\kappa_{\bf m}}{(|{\bf n}-{\bf
m}|^2+c^2)^{3/2}}
\end{equation}

where $\kappa_{\bf m}$ is some force coefficient, which
characterizes charge states of the upper chain's ions, and $|{\bf
n}-{\bf m}|$ is measured in units of $|\bf a|$. In this work we
choose $\kappa_{\bf
  m}$ as uniformly distributed random number in the interval of
$(0,1)$, so $\kappa_{\bf m}\neq \kappa_{\bf m'}$ and $\kappa_{\bf
  m}\in(0,1)$. For this reason the value of  $\kappa_{\bf m}$ changes
randomly in passing ${\bf m}$ from one site to another site.

\section{Results and discussion}

The possible ways of changing of the value of $k_{\bf m}$ are
illustrated in Fig.\ref{fig.2}. In turn density-displacement force
Eq.(\ref{5}) acquires a peculiarity of randomized (i.e. disordered)
electron-phonon interaction force. From now the force Eq.(\ref{5})
has two features: (i) it is still longer-ranged and descends as
$r^{-3}$, where $r$- is the distance between the electron and ion
under consideration, and (ii) it is randomized (disordered). In
Fig.\ref{fig.3} the plots of the dependence of the value of
density-displacement electron-phonon interaction force Eq.(\ref{5})
on site indexes $\bf m$ are presented \footnote{In figures
{\ref{fig.3}}, {\ref{fig.7}} and {\ref{fig.8}}
  blue-circle lines correspond to a regular lattice and are given for a comparison.}. As it is seen from the plots randomization of the electron-phonon interaction force changes the picture of the phenomenon drastically (qualitatively and quantitatively). Indeed, for a regular lattice we have $f_{{|\bf m|}}({\bf n})>f_{{|\bf m'|}}({\bf n})$ if ${|\bf m|}<{|\bf m'|}$. But for our case of the lattice with the disordered electron-phonon interaction the above relation is not always true i.e. on some sites $f_{{|\bf m|}}({\bf n})<f_{{|\bf m'|}}({\bf n})$ even if ${|\bf m|}<{|\bf m'|}$. This is the effect of considering disorder on the density-displacement type electron-phonon interaction. As a result of all parameters of the polaron will be affected by the disorder. In order to show this explicitly we calculate just some basic polaron's parameters like $E_p$- polaron shift, $g^2$- band-narrowing factor, $\gamma$- a numerical factor that depends on crystal struture and binds the polaron energy $E_p$ to band-narrowing  factor $g^2$ via relation $g^2=\gamma E_p/\hbar\omega$ and $\Phi({\bf n}-{\bf n'})$- whose gradient in corresponding direction is important to determine the mass of the polaron in a regular lattice.

\begin{figure}
\resizebox{0.45\textwidth}{!}{%
  \includegraphics{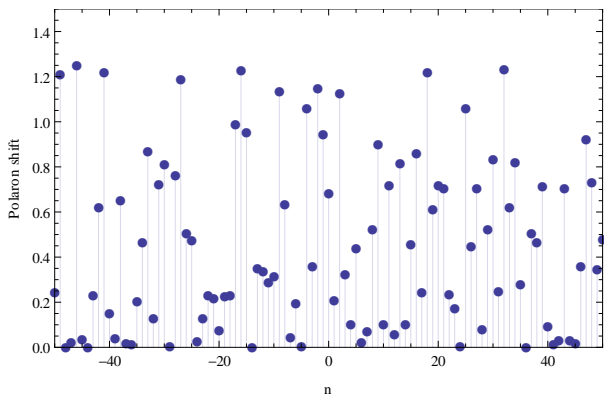}
}
\caption{The value of polaron shift $E_{P}$ at sites ${\bf n}$ due
to disordered electron-phonon interaction.}
\label{fig.4}       
\end{figure}

\begin{figure}
\resizebox{0.45\textwidth}{!}{%
  \includegraphics{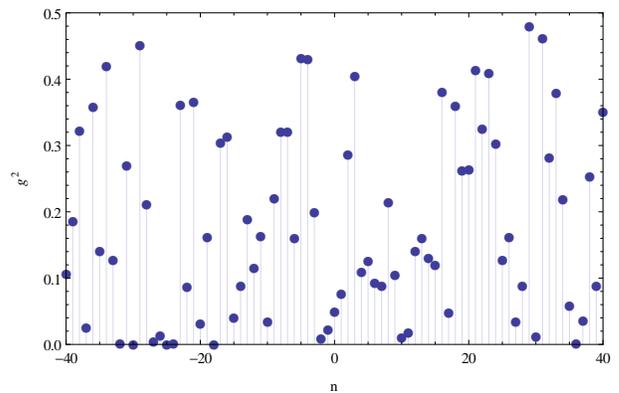}
}
\caption{The value of band-narrowing factor $g^{2}$ at sites ${\bf
n}$ due to disordered electron-phonon interaction.}
\label{fig.5}       
\end{figure}

\begin{figure}
\resizebox{0.45\textwidth}{!}{%
  \includegraphics{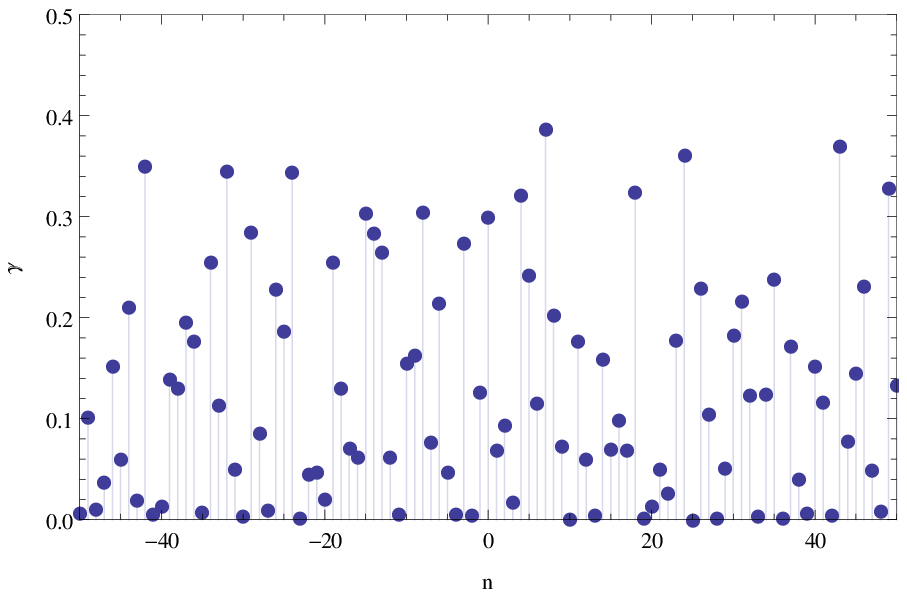}
}
\caption{The value of $\gamma$ at sites ${\bf n}$ due to disordered
electron-phonon interaction.}
\label{fig.6}       
\end{figure}

\begin{figure}
\resizebox{0.45\textwidth}{!}{%
  \includegraphics{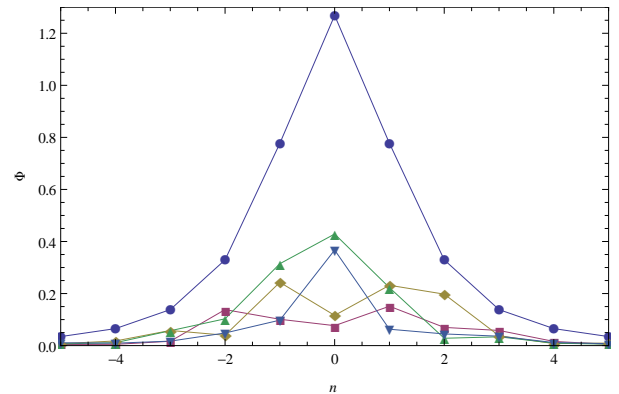}
}
\caption{The value of $\Phi ({\bf n})$ at sites ${\bf n}$ due to
disordered electron-phonon interaction.}
\label{fig.7}       
\end{figure}

In strong electron-phonon coupling limit and nonadiabatic regime one
uses the standard procedures such as Lang-Firsov transformation
\cite{lang-fir} that eliminates electron-phonon interaction term
(\ref{4}). Subsequent use of perturbation expansion of the
transformed Hamiltonian $H_e$ with respect to parameter
$\lambda^{-1}=2t/E_p$ up to the first order in the hopping integral
yields

\begin{equation}\label{6}
    E_{p}({\bf n})=\frac{1}{2M\omega^2}\sum_{{\bf m}}f^{2}_{{\bf m}}({\bf n}),
\end{equation}
\begin{equation}\label{7}
    g^2({\bf n},{\bf a})=\frac{1}{2M\hbar\omega^3}\sum_{{\bf m}}[f^2_{{\bf m}}({\bf n})-f_{{\bf m}}({\bf n})f_{{\bf m}}({\bf n}+{\bf a})],
\end{equation}
\begin{equation}\label{8}
    \gamma({\bf n},{\bf a})=1-\frac{\sum_{{\bf m}}f_{{\bf m}}({\bf n})\cdot f_{{\bf m}}({\bf n}+{\bf a})}{\sum_{{\bf m}} f_{{\bf m}}^2({\bf n})},
\end{equation}

and

\begin{equation}\label{9}
    \Phi({\bf n}-{\bf n'})=\sum_{\bf m}f_{{\bf m}}({\bf n})\cdot f_{{\bf m}}({\bf n'})
\end{equation}

Though these parameters were obtained by perturbation theory under
certain assumptions we accept them as starting point expressions for
the discussion of the influence of disordered electron-phonon
interaction on polaron's parameters. In Fig.\ref{fig.4},
Fig.\ref{fig.5} and Fig.\ref{fig.6} we presented the plots of values
of  $E_p(\bf n)$, $g^2({\bf n},{\bf a})$ and $\gamma({\bf n},{\bf
a})$, respectively, on different sites ${\bf n}$. As it is seen from
the presented date all these parameters are randomized too.
Depending on disordered electron-phonon interaction polaronic effect
might be enhanced or diminished. The values of $\Phi({\bf n})$ on
different sites ${\bf n}$ are given in Fig.\ref{fig.7}.
Randomization undergoes also polaron's hopping integral from one
site to on another site of the lattice. Within the used framework
one can estimate the value of hopping integral from ratio
$\tilde{t}/t\simeq \exp(g^2)$ and plot its values for different
disorders, Fig.\ref{fig.8}. One can observe the same tendency for
hopping integral. Findings indicate that disorder changes a picture
of polaron formation, and it is appears that all polaron's
parameters are site dependent i.e disordered. Something similar
ideas, namely about site dependence of polaron energy (shift) and
local electron-phonon interaction's coupling constant on ${\bf n}$,
can be found in \cite{hopping-cond-in-sol}.

\begin{figure}
\resizebox{0.45\textwidth}{!}{%
  \includegraphics{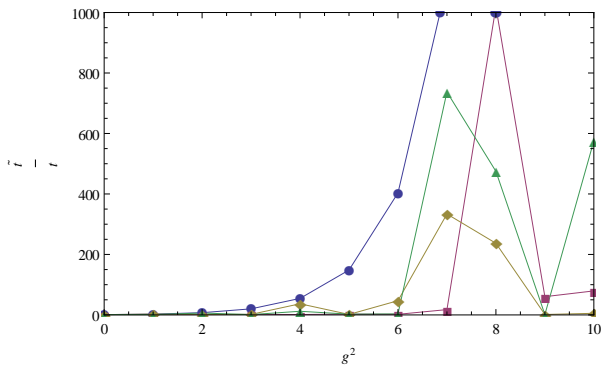}
}
\caption{ The value of renormalized hopping integral versus $g^2$
calculated taking into account disordered character of
electron-phonon interaction.}
\label{fig.8}       
\end{figure}

The above results were obtained under assumption that the values of
$\kappa$ are uniformly distributed within the interval (0,1). But
our model can be straightforwardly generalized to the case when the
values of $\kappa$ belong to an arbitrary interval $(p,q)$ and is
defined  within it by some function of distribution (Gaussian,
Poisson, Bernoulli and so on). There are a great variety of options
for simulation here. In our model it is also possible to simulate
string like structures and study a polaron formation in such
structures. For this purpose it is sufficient to select a required
set of values of $\kappa_{\bf m}$, for example $\kappa_{\bf
m}=(\ldots,0,0,1,1,1,1,0,0,1,1,1,1,\ldots)$. There are also other
possibilities.

Another question that we would like to discuss here concerns the
size of the polaron in the lattice where electron-phonon interaction
is disordered. As for the Hamiltonian (\ref{1}) there are two points
of view on this matter. Ref.\cite{alekor} treats the polaron of the
model (\ref{1}) as \textit{small} Fr\"{o}hlich polaron, while
Ref.\cite{flw} treats the polaron of the model (\ref{1}) as
\textit{large} Holstein polaron. Here we follow the definition in
which small polaron is formed at $t/E_p\ll 1$ and large polaron is
formed at $t/E_p\gg 1$ \cite{hopping-cond-in-sol}. Since in our
model the ratio $t/E_p$ fluctuates from site to site one can not say
clearly about size of polaron even it formed by long-rang
electron-phonon interaction (\ref{5}). For some sites the condition
of small (large) polaron formation is satisfied, while for other
sites it is ruled out. Then one may say that in the systems with
disordered electron-phonon interaction polaron's size also modulated
by the disordered electron-phonon interaction. It seems that such
property of polarons in disordered structures is common since in
Ref.\cite{mishchenko-nagaosa-etal} a mixed behavior typical for
large (away impurity) and small (near impurity) polarons was also
expected for a crystal with impurities. The results obtained here
suggest further study of polaron formation in disordered structures
taking into an account not only on-diagonal and off-diagonal aspects
of the disorder, but also the disordered nature of the
electron-phonon interaction as well.

\section{Conclusion}
In conclusion we have studied one dimensional lattice within the
framework of the extended Holstein model. Considering the case when
charge fluctuations can occur in the system we have applied the
extended Holstein model to study polaron formation in the system. In
doing this we introduce a disordered density-displacement type
electron-phonon interaction by randomizing force constant that
describes charge state of the ions. We have calculated the basic
parameters of polaron with the disordered electron-phonon
interaction and have found that all parameters of polaron are
drastically affected by the disordered electron-phonon interaction.
It turns out that all parameters of polaron are site dependent. This
is true with respect to ratio $t/E_p$ that determines a polaron
size. Then one can not say definitely about size of polaron in such
system. The size of the polaron depends on its position and from the
disorder seen by polaron from this location. So, the size of the
polaron is modulated (disordered) i.e. site dependent in the system
with disordered electron-phonon interaction.

\begin{acknowledgments}
Author (B.Ya.) would like to thank the Abdus Salam International
Centre for Theoretical Physics (Trieste, Italy) for its hospitality
throughout a visit during which this manuscript was written. Author
also appreciates a fruitful discussions with Prof. Vladimir
E.Kravtsov and Dr. Rachidi A.A.Yessoufou.
\end{acknowledgments}

\end{document}